%% file: main.tex
\documentclass[a4paper,11pt]{article}
\pdfoutput=1 % if your are submitting a pdflatex (i.e. if you have
\usepackage[comma,square,numbers,sort&compress]{natbib}
\usepackage{jinstpub} % for details on the use of the package, please
\usepackage{lineno}
\usepackage{subfigure}

\usepackage{float}
\usepackage{xcolor}
\usepackage{listings}
\definecolor{codegreen}{rgb}{0,0.6,0}
\definecolor{gray}{rgb}{0.5,0.5,0.5}
\definecolor{codepurple}{rgb}{0.58,0,0.82}

\lstdefinestyle{myCPPStyle}{
  language=C++,
  aboveskip=3mm,
  belowskip=3mm,
  showstringspaces=false,
  columns=flexible,
  basicstyle={\small\ttfamily},
  numbers=none,
  numberstyle=\scriptsize\color{gray},
  keywordstyle=\color{blue},
  commentstyle=\color{codegreen},
  stringstyle=\color{magenta},
  breaklines=true,
  breakatwhitespace=true,
  tabsize=2
}

\usepackage{xspace}
\input{commands}

\title{\boldmath Using Machine Learning for Particle Identification in ALICE}

\author[a,1]{{\L}ukasz Kamil Graczykowski,\note{Corresponding author.}}
\author[b]{Monika Jakubowska,}
\author[c]{Kamil Rafa{\l} Deja}
\author[a]{and Maja Kabus}

\affiliation[a]{Faculty of Physics, Warsaw University of Technology\\Koszykowa 75, 00-662 Warsaw, Poland}
\affiliation[b]{Faculty of Electrical Engineering, Warsaw University of Technology\\Pl. Politechniki 1, 00-661 Warsaw, Poland}
\affiliation[c]{Faculty of Electronics and Information Technology, Warsaw University of Technology\\Nowowiejska 15/19, 00-665 Warsaw, Poland }

\emailAdd{lukasz.graczykowski@pw.edu.pl}

\abstract{Particle identification (PID)
is one of the main strengths of the ALICE experiment at the LHC. It is a crucial ingredient for detailed studies of the strongly interacting matter formed in ultrarelativistic heavy-ion collisions. ALICE provides PID information via various experimental techniques, allowing for the identification of particles over a broad momentum range (from around 100~\MeVc to around 50~\GeVc). The main challenge is how to combine the information from various detectors effectively. Therefore, PID represents a model classification problem, which can be addressed using Machine Learning (ML) solutions. Moreover, the complexity of the detector and richness of the detection techniques make PID an interesting area of research also for the computer science community. In this work, we show the current status of the ML approach to PID in ALICE. We discuss the preliminary work with the Random Forest approach for the LHC Run 2 and a more advanced solution based on Domain Adaptation Neural Networks, including a proposal for its future implementation within the ALICE computing software for the upcoming LHC Run 3.}


\collaboration[c]{on behalf of the ALICE collaboration}

\proceeding{AI4EIC-Exp -- Experimental Applications of Artificial Intelligence for the Electron-Ion Collider\\
7-10 September 2021\\
Brookhaven National Laboratory, Upton (NY), USA}

\begin{document}
\maketitle
\flushbottom

\section{Introduction}
\input{ch1_intro_pid}

\section{Preliminary work with Random Forest in LHC Run 2}
\label{sec:Run2}
\input{ch2_pid_run2}

\section{ALICE computing framework in LHC Run 3}
\input{ch3_o2_analysis_framework}

\section{The proposed solution for particle identification in LHC Run 3}
\input{ch4_pid_run3}

\input{ch5_conclusions}

\acknowledgments
Research was funded by POB HEP of Warsaw University of Technology within the Excellence Initiative: Research University (IDUB) program as well as by the Polish National Science Center grants no. UMO-2018/31/N/ST6/02374, UMO-2017/27/B/ST2/01947, UMO-2016/22/M/ST2/00176 and the Polish Ministry of Education and Science under agreement no. 2022/WK/01.

\bibliographystyle{unsrt}
\bibliography{bibliography}

\end{document}

%% file: commands.tex
\newcommand{\pp}           {pp\xspace}

\newcommand{\PbPb}         {\mbox{Pb--Pb}\xspace}

\newcommand{\pPb}          {\mbox{p--Pb}\xspace}

\newcommand{\s}            {\ensuremath{\sqrt{s}}\xspace}

\newcommand{\pt}           {\ensuremath{p_{\rm T}}\xspace}

\newcommand{\dEdx}         {\ensuremath{\textrm{d}E/\textrm{d}x}\xspace}

\newcommand{\nineH}        {$\sqrt{s}~=~0.9$~Te\kern-.1emV\xspace}
\newcommand{\seven}        {$\sqrt{s}~=~7$~Te\kern-.1emV\xspace}
\newcommand{\twoH}         {$\sqrt{s}~=~0.2$~Te\kern-.1emV\xspace}
\newcommand{\twosevensix}  {$\sqrt{s}~=~2.76$~Te\kern-.1emV\xspace}
\newcommand{\five}         {$\sqrt{s}~=~5.02$~Te\kern-.1emV\xspace}
\newcommand{\twosevensixnn}{$\sqrt{s_{\mathrm{NN}}}~=~2.76$~Te\kern-.1emV\xspace}
\newcommand{\fivenn}       {$\sqrt{s_{\mathrm{NN}}}~=~5.02$~Te\kern-.1emV\xspace}

\newcommand{\GeVc}         {Ge\kern-.1emV/$c$\xspace}
\newcommand{\MeVc}         {Me\kern-.1emV/$c$\xspace}
\newcommand{\TeV}          {Te\kern-.1emV\xspace}
\newcommand{\GeV}          {Ge\kern-.1emV\xspace}
\newcommand{\MeV}          {Me\kern-.1emV\xspace}
\newcommand{\GeVmass}      {Ge\kern-.2emV/$c^2$\xspace}
\newcommand{\MeVmass}      {Me\kern-.2emV/$c^2$\xspace}

\newcommand{\osq}         {\ensuremath{\rm{O}^2}\xspace}

%% file: ch1_intro_pid.tex
ALICE (A Large Ion Collider Experiment)~\cite{Aamodt:2008zz} is one of the four big detectors located at the Large Hadron Collider~(LHC) at CERN~\cite{Evans:2008zzb}. Its main goal is widening our understanding of the physics of ultrarelativistic heavy-ion collisions and measuring the properties of the quark--gluon plasma (QGP), a deconfined state of quarks and gluons, theorized to exist in the early Universe~\cite{Shuryak:1978ij,Adams:2005dq}. The detailed characterization of QGP requires utilizing several observables, which are calculated from the lead nuclei collision (\PbPb) data, systematically compared with the proton-proton (\pp) and proton-lead (\pPb) collision data, and interpreted through various theoretical models~\cite{Foka:2016vta,Foka:2016zdb}. 

The apparatus is composed of various detector systems that use a number of techniques to measure signals from the particles emitted in such collisions. The so-called ``central barrel'' surrounds the collision point and consists of a large solenoid magnet that generates a uniform magnetic field of up to 0.5~T along the beam direction. Inside the magnet, a set of detectors is located surrounding the beam axis radially. Additional detectors, such as the muon spectrometer, are located outside the central barrel in the forward beam direction. One of the main aspects of ALICE, differentiating it from other LHC experiments and a pre-requirement for detailed QGP studies, is its particle identification (PID) capability, {\it i.e.}, the discrimination power for different particle species produced in a collision. Due to a number of available detectors which operate concurrently, the signal separation for various types of particles is possible over a wide range of momentum, from just around 100~\MeVc up to around 10~\GeVc.
For the purpose of the work presented in this study, three detectors are most relevant. For the track reconstruction and hadron PID, the information provided by the Inner Tracking System (ITS)~\cite{ALICE:1999cls}, the Time Projection Chamber (TPC)~\cite{ALICE:2000jwd,Garabatos:2004iv}, and the Time-of-Flight detector (TOF)~\cite{CERN-LHCC-2000-012,Antonioli:2003cw,Akindinov:2013tea} is used. For a detailed description of the ALICE detector performance, we refer the reader to Ref.~\cite{ALICE:2014sbx}. 

The main task of PID is to provide high purity samples of particles of a given type required by the analyzer conducting a specific analysis. In the most traditional approach, particles are selected by applying so-called ``cuts'' applied to the reconstructed features obtained from the detector response,
rejecting the particles not meeting the specified criterion. Such an approach is justified when the separation between various particle species is significantly large. However, when the feature distributions associated with the particle species begin to overlap, the process of combining the information from multiple detectors becomes non-trivial, and the ``trial and error'' approach based on the intuition and experience of the analyzer becomes not optimal. This leads to lowered PID efficiency and limits the statistical significance of the final data analysis. These shortcomings can be addressed with more advanced classification methods. One particular example is the Bayesian approach, which has proven successful in the ALICE experiment~\cite{ALICE:2016zzl}. However, other solutions, i.e., those based on Machine Learning (ML), also have a potentially significant impact. In particular, in our preliminary studies done with the LHC Run 2 data in Ref.~\cite{trzcinski2018using}, we show that classifiers based on Random Forest~\cite{ho1995random} improve the separation between various particle species significantly. Nonetheless, one of the main limitations of that work was not taking into account any discrepancies between the Monte Carlo (MC) and experimental data.

This paper presents a status of ongoing activity to deliver a comprehensive PID framework for LHC Run 3, with a new computing framework and data format (called \osq) being deployed~\cite{tdr}. Based on what we have learned in the preliminary work for LHC Run 2, we propose a novel and more advanced solution based on Domain-Adversarial Training of Neural Networks~\cite{ganin2016domain}, which also addresses the MC and experimental data misalignment.

%% file: ch2_pid_run2.tex
For data gathered throughout the LHC Run 2, several attempts to introduce ML-based PID strategies were made. For example, in~\cite{trzcinski2018using}, we proposed a method based on a Random Forest~\cite{ho1995random} algorithm and showed the results for kaon selection. This technique is based on the idea of ensemble methods, where a group of weak decision trees is used to produce the final output, which corresponds to the class predicted by most of the individual classifiers. Each tree is trained with only a subset of all available parameters and a subset of available data examples to increase their variance. In~\cite{trzcinski2018using}, the tree generation method based on the Gini index was used, which is defined as the probability of the wrong classification while using only a given attribute. Proposed experiments indicate that thanks to incorporating additional track-related attributes, the ML-based PID provides much higher efficiency and purity for the selected particles than standard methods. 
In particular, in~\cite{trzcinski2018using}, we showed a comparison of purity and efficiency as a function of \pt for kaons selected with the traditional method ($n_{\rm \sigma,TCP}<2$ for \pt$\leq0.5$~\GeVc and $\sqrt{n_{\rm \sigma,TCP}^2+n_{\rm\sigma,TOF}^2}<2$ for \pt$>0.5$~\GeVc, where for a given particle measured in a detector, $n_{\sigma}$ is the number of standard deviations from the expected value), and using the Random Forest classifier for PYTHIA~6.4~\cite{Sjostrand:2006za,Sjostrand:2019zhc} (Perugia-0 tune~\cite{Skands:2010ak}) simulated MC \pp data at a collision energy of $\s=7$~\TeV. In this study on simulated data, the Random Forest classifier outperformed the traditional cut-based selection.

Even though the proposed method provided significant improvement in PID performance, its incorporation in the~\emph{AliRoot}~\cite{Brun:2003vw} computing framework has turned out to be very difficult. The reason is the communication between the Python environment, in which the ML classifiers were prepared, and C++, which is the basis of ROOT~\cite{BRUN199781} and \emph{AliRoot}. Each track needs to be propagated to the classifier to receive a response with probabilities assigned to it for each particle species. We considered several attempts to propagate tracks on a track-by-track basis or propagate chunks of tracks combined from single or multiple events to the classifier. However, since the processes running the analysis in \emph{AliRoot} on the GRID cannot be parallelized, this was a significant limitation, and the fastest solution had a processing time more than 3 times slower than the traditional PID. However, with the end of the LHC Run 2 data-taking period and preparations ongoing for the upcoming LHC Run 3 with the new ALICE computing software more open to external ML tools, it was decided not to continue attempting to solve those problems related to \emph{AliRoot}.

%% file: ch3_o2_analysis_framework.tex
Currently, the LHC is being upgraded before the next data-taking period, LHC Run 3, which will start in the spring of 2022. The ALICE experiment will need to cope with a 100 times bigger data-taking rate than the LHC Run 2. Therefore, the upgrade includes replacing some detectors like ITS and TPC and the data acquisition software. Moreover, the~new detector will continuously collect the~data instead of using a triggered readout, switching the data readout on and off each time a collision is expected~\cite{ALICE:2012dtf}.

\begin{figure}[ht]
    \centering
    \includegraphics[width=\textwidth]{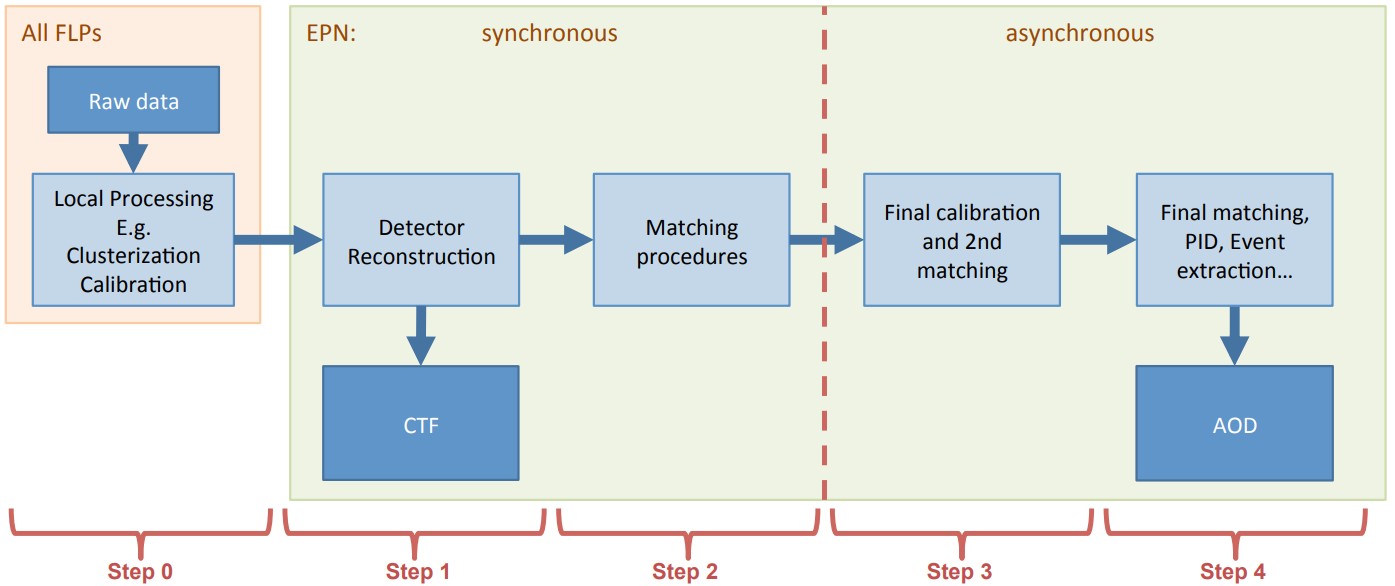}
    \caption{Reconstruction and calibration data flow (FLP -- First Level Processors, EPN -- Event Processing Nodes, CTF -- Continuous Time Frames, AOD -- Analysis Object Data). Figure from~\cite{tdr}.}
    \label{fig:o2-ana-flow}
\end{figure}

The new ALICE software called the ALICE Online-Offline (\osq) framework is being developed anew to allow for high throughput, hardware acceleration, and more concurrency~\cite{tdr,ALICE:2014lor}. Data reconstruction will happen in two main stages, the synchronous (online) and the asynchronous (offline) stage~\cite{tdr}. Synchronous processing encompasses fast, rough calibration and reconstruction that substantially reduces the data volume at an early stage. The data is then compressed and stored for the asynchronous stage at which final calibration is applied to achieve permanently stored high-quality output. Therefore, a considerable speed-up will be conducted by performing detector calibration and data reconstruction (traditionally offline/asynchronous) also partially online (at the synchronous stage), concurrently with data-taking. Figure~\ref{fig:o2-ana-flow} presents a simplified scheme of the reconstruction and calibration data flow. After the final calibration, particle identification will take part at the asynchronous stage.

\subsection{$\mathbf{ \rm O^2}$ Analysis Framework}

One of the main parts of the \osq software is the Analysis Framework, which encompasses tools for physics data analysis~\cite{Alkin:2021mfo}. It provides efficient data processing with an object-like, user-friendly API, making it easier for the users to migrate from the old software framework and taking implicit advantage of vectorization and parallelization. The solution is based on Apache Arrow for data representation, allowing integration with other tools like Python Pandas. The Analysis Framework is integrated with the \osq Data Processing Layer (DPL) that encompasses the whole data flow from data acquisition to its analysis.

The event data is reduced as much as possible to reduce disk space requirements and speed-up retrieval. It follows the LHC Run 2 AOD (Analysis Object Data) design but without any quantities that can be calculated on the fly. The data model no longer consists of objects, but it is represented as flat tables similar to relational databases and stored as flat ROOT trees. This will save the costs due to serialization during message passing, enable vectorized processing, and allow for more efficient use of the shared memory backend of FairMQ (Message Queuing Library and Framework)~\cite{fairmq}. Nevertheless, the analysis output and histograms will be serialized as objects with ROOT.

An analysis is performed in a workflow, a sequence of tasks. Each task is a struct that needs to contain at least the \lstinline{process()} method. The users can subscribe to specific tables by passing them as parameters to \lstinline{process()}. Other task struct members can define filters and partitions to apply or new tables and histograms to be produced.

Listing~\ref{lst:filtering} shows a sample task. The tracks are processed from one collision at a time, and they are filtered according to their $\eta$. Inside the \lstinline{process()} method, a 2D histogram is filled with $\eta$ and $\phi$ of the filtered tracks and then tracks with positive and negative $\phi$ are processed separately.

\begin{lstlisting}[language=C++, captionpos=b,
basicstyle={\scriptsize\ttfamily},
caption={An example of filtering and partitioning tracks.},
label={lst:filtering}]
struct FilterTask {
  using myTracks = soa::Filtered<aod::Tracks>
  Filter etafilter = (aod::track::eta < 1.0f) && (aod::track::eta > -1.0f);
  Partition<myTracks> leftPhi = aod::track::phiraw < 0.0f;
  Partition<myTracks> rightPhi = aod::track::phiraw >= 0.0f;
  OutputObj<TH2F> etaphiH{TH2F("etaphi", "etaphi", 100, 0., 2. * M_PI, 102, -2.01, 2.01)};

  void process(aod::Collision const& collision, myTracks const& tracks)
  {
    for(auto& track : tracks) {
      etaphiH->Fill(track.phi(), track.eta());
    }
    for (auto& track : leftPhi) { /*Processing filtered tracks from the first partition*/ }
    for (auto& track : rightPhi) { /*Processing filtered tracks from the second partition*/ }
  }
};
\end{lstlisting}

Since the results of particle identification will be directly used by the analyzers, the PID framework needs to efficiently take the inputs from the \osq analysis tasks and pass the calculated particle codes back to user-defined analysis tasks.

%% file: ch4_pid_run3.tex
To improve particle identification performance, we propose to implement a classifier based on a multilayered perceptron. We train it with data from corresponding MC simulations. To simplify integration and allow the selection of individual particles independently, we propose to train one model with a binary classification objective for each particle type.

\subsection{Domain Adaptation}

In analysis, PID is used to select particles of desired types in both real experimental data and Monte Carlo simulations. However, recorded signals in physical detectors can differ from those produced in simulations. Therefore, standard PID methods rely on partially automated processes for data domains alignment. For instance, one of ALICE's approaches consists of using the so-called ``tune on data''. This procedure is based on generating a random detector signal based on a parametrization obtained from data. The resulting distributions of Monte Carlo and experimental data should be equivalent, and only statistical fluctuations are present. 

To circumvent the limitations of standard data alignment methods, we propose combining it with particle identification stating it as a known problem of classification with unsupervised domain adaptation. The main idea of this technique is to learn the discrepancies between two data domains, that is, the labeled source domain (in our case simulation data) and the unlabelled target one (in our case experimental data), and translate those to a single hyperspace, where the differences between domains are no longer visible. Classifiers trained on top of features located in combined latent space should have similar performance on both MC simulated and experimental data. Nevertheless, in such a scenario, simulation data is crucial to learn how to distinguish different particles based on aligned representations.
The domain-adaptation technique is widely used in natural language processing~\cite{blitzer2007biographies,glorot2011domain} and computer vision~\cite{gopalan2011domain,fernando2013unsupervised}. In the domain of high-energy-physics, its application is limited only to preliminary studies of jet classification~\cite{walter2018domain}. In this work, the authors present that this method can improve the quality of automatic jet tagging on real experimental data.

Our initial experiments with Domain Adversarial Neural Networks (DANN)~\cite{ganin2016domain} show that this technique improves the classification of particles in experimental data. The main idea behind this method is to build a system composed of three neural networks. The goal of the feature mapping network is to map original input into domain invariant features. Those features serve as an input to the standard particle classifier that outputs the particle type. Additionally, the last model, known as domain classifier, enforces domain invariance of extracted features through adversarial training procedure. The training of the model is divided into two steps. First, on top of current features from the feature mapping network, the domain classifier is trained independently to classify domain labels -- whether data come from a real or a simulation source. Then, the domain classifier is being frozen so that the particle classifier and the feature mapper can be trained jointly to predict accurate particle types while fouling the domain classifier at the same time. With this approach, the feature mapper's weights are updated with a gradient from the particle classifier and reversed gradient from the domain classifier. Training a domain-adaptation-based classifier is more complex than the classical neural model. However, the application performance of those two methods is similar and depends on the complexity of the classifier and feature extractor.

\begin{figure}[t]
    \centering
    \includegraphics[width=0.8\textwidth]{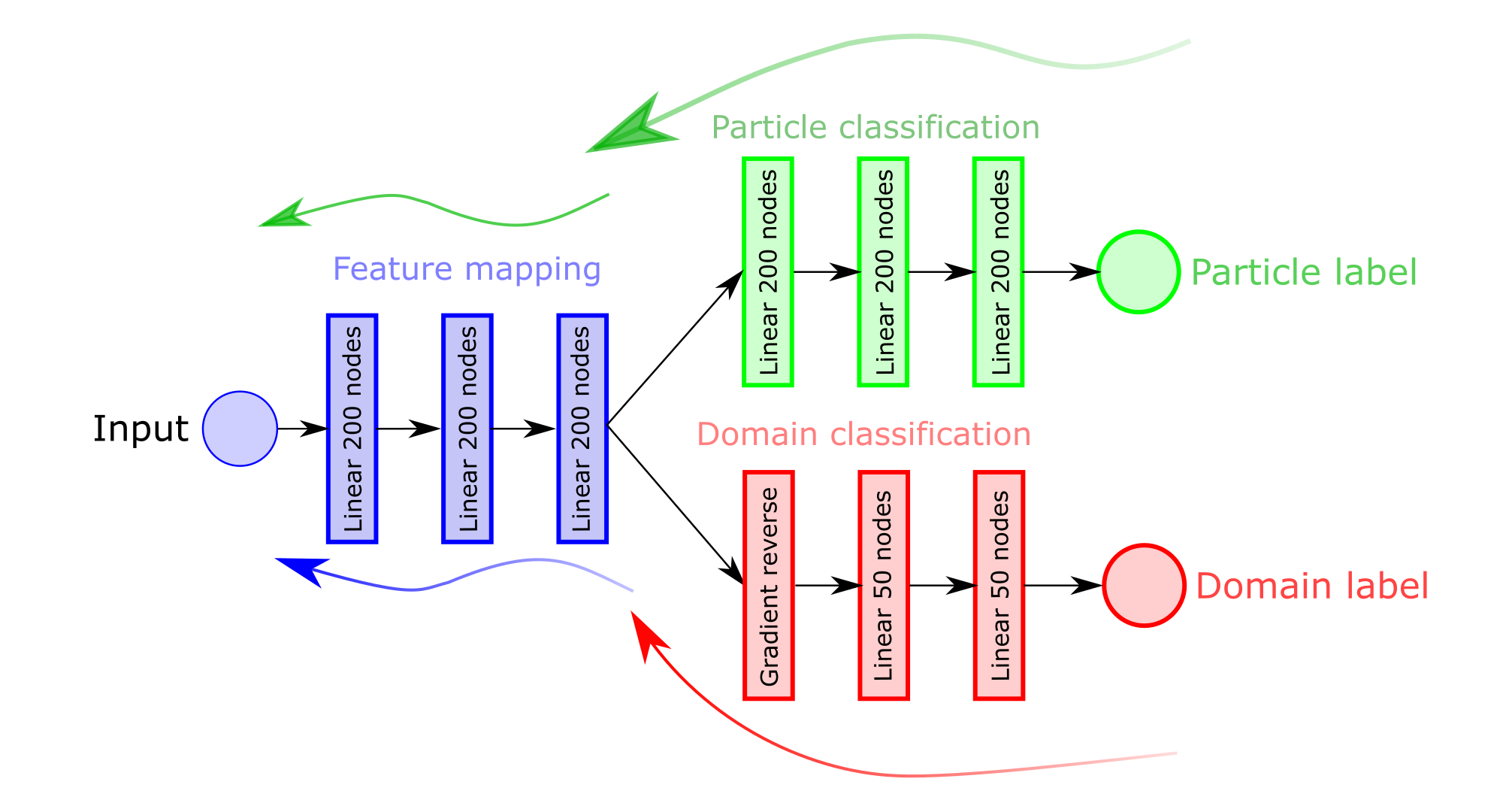}
    \caption{Architecture of DANN composed of three models: feature mapper, particle classifier and domain classifier.}
    \label{fig:pid-dann}
\end{figure}

\begin{figure}[t]
 		\centering
 		\subfigure{\includegraphics[width=.49\linewidth]{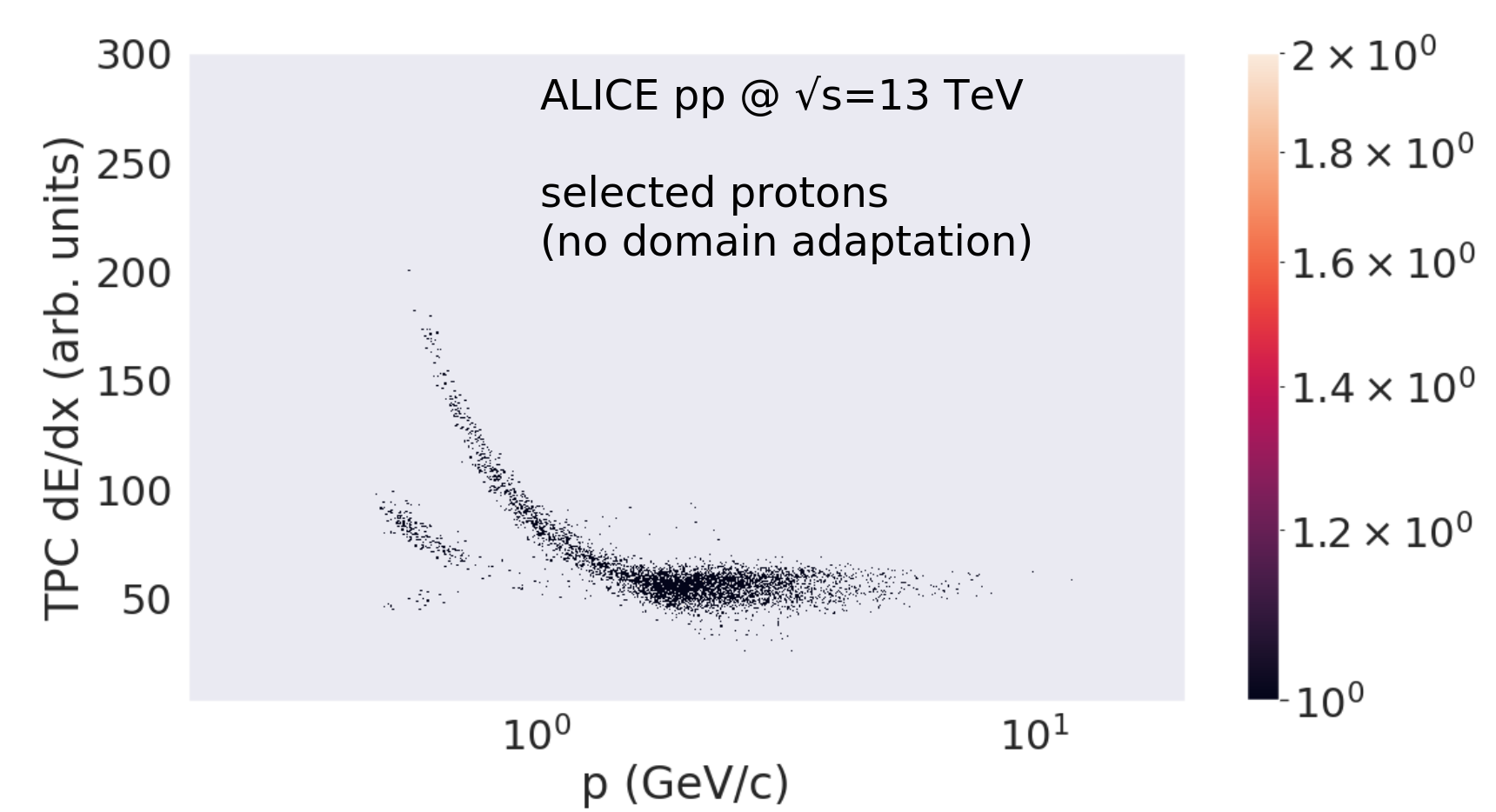}}
		\subfigure{\includegraphics[width=.49\linewidth]{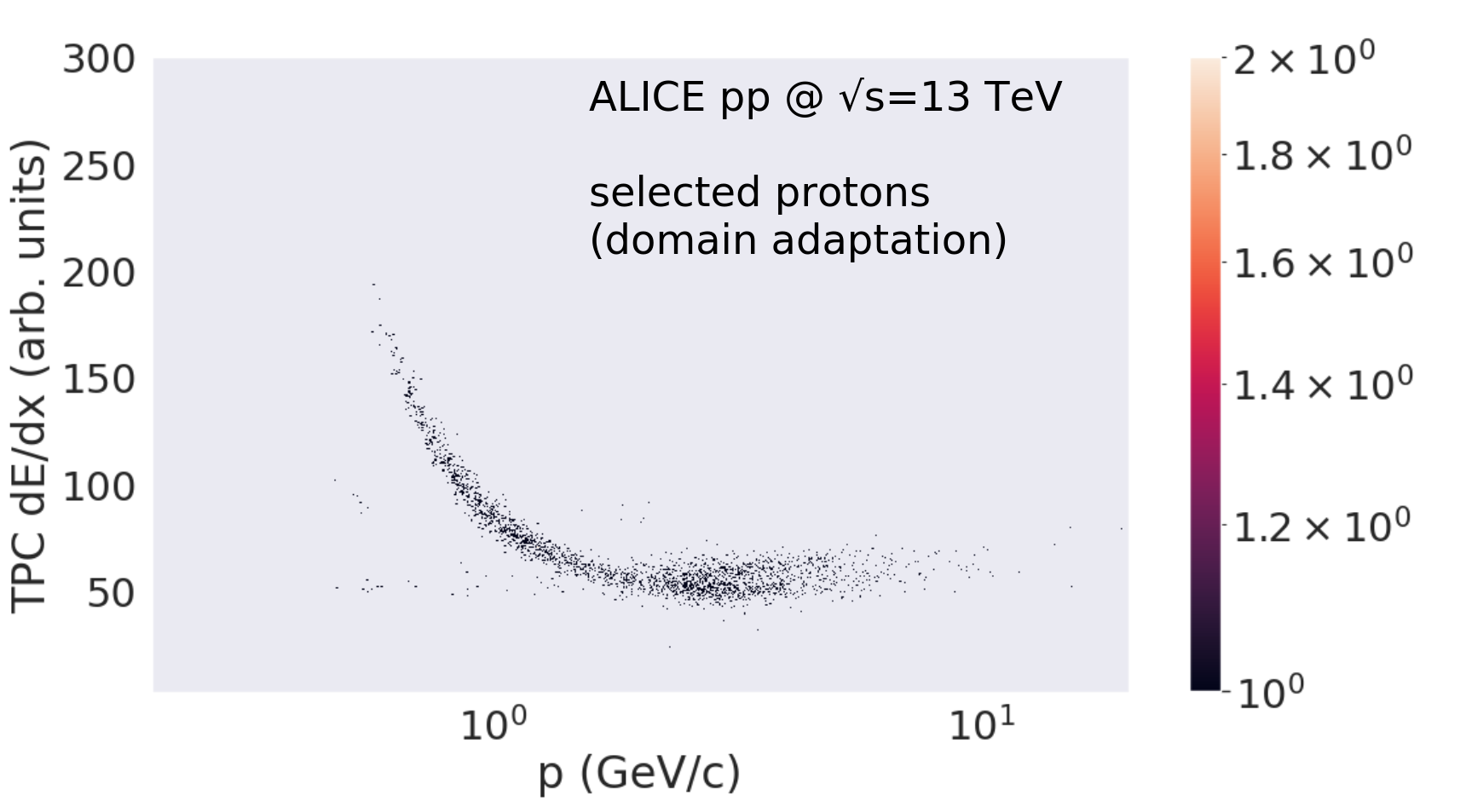}}
\caption{\label{fig:da_results}Preliminary result of DANN PID for the TPC detector signal (\dEdx) as a function of particle momentum for particles identified as protons without domain adaptation (left) and with domain adaptation (right).}
\end{figure}

Figure~\ref{fig:da_results} presents preliminary results of the proposed DANN model for pion identification in \pp data at $\s=13$~\TeV from the LHC Run 2 period. The training dataset was a corresponding MC simulation with PYTHIA~8 Monash tune~\cite{Skands:2014pea}.

Our preliminary results of classification applied with the proposed model indicate that domain adaptation improves experimental data classification. The enhancement is visible as a reduction of contamination outside the proton band in the energy loss signal of protons classified with DANN in Fig.~\ref{fig:pid-dann}. However, detailed benchmarks will have to be done in the future after the final architecture of the model is achieved in order to support that hypothesis. 

\subsection{Implementation of the Machine Learning PID analyses with the $\rm O^2$ framework}
\begin{figure}[ht]
    \centering
    \includegraphics[width=\textwidth]{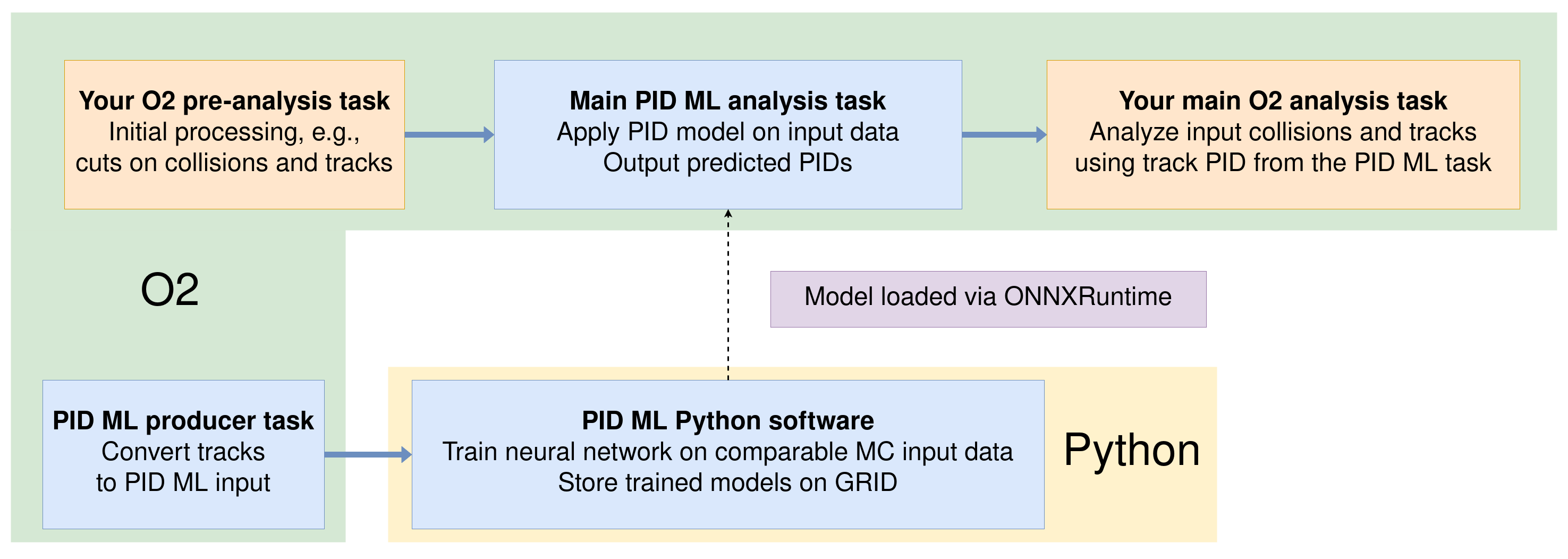}
    \caption{An initial scheme of the particle identification workflow in \osq. More information about the workflow topology in the \osq Analysis Framework can be found in~\cite{Alkin:2021mfo}.}
    \label{fig:pid-o2-scheme}
\end{figure}

Besides developing and improving the neural network models, the new particle identification framework needs to be integrated into the more extensive analysis software. Figure~\ref{fig:pid-o2-scheme} depicts an initial, tentative design of the PID workflow, which will be further tested and updated.

The current developments make use of the ONNX (Open Neural Network Exchange) standard~\cite{onnx}, which defines a common file format for storing machine learning models developed in various frameworks such as Tensorflow and PyTorch. Additionally, the ONNX Runtime~\cite{onnxruntime} library enables ONNX models to be used in different programming languages such as Python and C++ with a simple API. ONNX and ONNX Runtime are also tested by other machine learning projects at ALICE, and the solution provided for PID ML will be an example base for the other projects.

Firstly, it is assumed that the reconstructed collision data will be available in AOD files. Separately, different trained and untrained neural networks will be stored in their own repository in ONNX format. PID analysis tasks in the \osq Analysis Framework will comprise the PID ML API available to users. Once an analyzer requests a PID task in their workflow, a PID producer task will be first attached. This task will read appropriate collision data and distill the variables needed for the identification task so that the further processing will operate on a skimmed, much smaller table.

Then, the framework will determine whether a matching neural network is available. If the network requires training, the training task is triggered. The training process itself might be performed on the C++ or Python side -- this also depends on ONNX Runtime limitations. In the case of Python, the data table will be converted from stored ROOT trees to Python pandas.

Finally, the ready model is passed back to the Analysis Framework, and the inference task is triggered. Currently, the inference is implemented both in Python and in C++ (with ONNX Runtime). The tests presented in this work were performed in Python. The inference results will then be passed to the user analysis task.

Initially, a multi-output neural network was considered, but it was rejected in favor of having different models for recognizing different particle kinds. In this way, the development can be concentrated on the most commonly used particle species.

%% file: ch5_conclusions.tex
\section{Future works}
This work presents the current status of the development of particle identification techniques based on machine learning. Several main concerns still need to be addressed for those methods to work in practice.

To assess the precision of measurements based on ML-PID, we have to estimate the systematic uncertainties of machine learning models. While this is still an open research question, several methods approximate different sources of systematic uncertainties. In~\cite{jha2019impact}, the authors demonstrate how dataset uncertainties can affect machine learning model predictions. Among the works related specifically to high-energy physics, in~\cite{ghosh2021uncertainty}, the authors propose a model where classifiers are aware of systematic uncertainties of input parameters. This approach improves the model's sensitivity. At the same time, in~\cite{englert2019machine}, the authors use adversarial training for this purpose.

In our solution, we plan first to include experiments analogous to~\cite{jha2019impact} measuring how training dataset selection might affect the model's performance.
Next, we intend to follow the developments described in~\cite{gal2016dropout} where the authors show that dropout -- a standard method for reducing overfitting in neural networks -- might be used to approximate Bayesian uncertainty in deep Gaussian processes. A method for calibrating Bayesian uncertainties is presented also in~\cite{kuleshov2018accurate}.

\section{Conclusions}
Identification of various particle species is one of the strengths of ALICE and a crucial ingredient of many analyses studying the QGP properties from heavy-ion data analysis. While the typical PID methods are based on more or less arbitrary cut-off values selected by the analyzers, improvement can be achieved with ML-based approaches. PID can be seen as a multi-class classification problem presenting the challenges described in the text, which can be of interest to other communities. This fosters multidisciplinary approaches like the ones we proposed. The preliminary work from the LHC Run 2 with the Random Forest approach already shows that using ML techniques for PID improves the purity of selected samples while maintaining high efficiency. However, one of the essential ingredients is the misalignment between the Monte Carlo training data and the experimental data. A solution to that problem is domain adaptation via the DANN networks. The first results obtained with DANN networks show that this approach is promising. The implementation in the new ALICE software framework (the \osq via the ONNX environment) should also be feasible without the drawbacks present in the preliminary work for LHC Run 2 data and the previous \emph{AliRoot} framework.